\documentclass[prc,preprint,showpacs,preprintnumbers,amsmath,amssymb]{revtex4}
\usepackage{graphicx}
\newcommand{\beq}{\begin{equation}}
\newcommand{\eeq}{\end{equation}}
\newcommand{\beqa}{\begin{eqnarray}}
\newcommand{\eeqa}{\end{eqnarray}}

\newcommand{\Sigs}{\Sigma_{\mathrm s} }
\newcommand{\Sigv}{\Sigma_{\mathrm v} }
\newcommand{\Sigo}{\Sigma_{\mathrm o} }
\newcommand{\kf}{k_{\mathrm F} }
\newcommand{\kfn}{k_{\mathrm Fn} }
\newcommand{\kfp}{k_{\mathrm Fp} }
\newcommand{\bfgamma}{\mbox{\boldmath$\gamma$\unboldmath}}
\newcommand{\veck}{\textbf{k}}

\begin{document}
\preprint{}
\title{Momentum, Density, and Isospin dependence of the Symmetric and Asymmetric Nuclear Matter Properties.}
\author{E. N. E. van Dalen}
\author{C. Fuchs}
\author{Amand Faessler}
\affiliation{Institut
f$\ddot{\textrm{u}}$r Theoretische Physik, Universit$\ddot{\textrm{a}}$t
T$\ddot{\textrm{u}}$bingen,
Auf der Morgenstelle 14, D-72076 T$\ddot{\textrm{u}}$bingen, Germany}
\begin{abstract}
Properties of symmetric and asymmetric nuclear matter have been investigated in the relativistic Dirac-Brueckner-Hartree-Fock approach based on projection techniques using the Bonn A potential.
The momentum, density, and isospin dependence of the optical potentials and nucleon effective masses are studied.
It turns out that the isovector optical potential depends sensitively on density and momentum, but is almost insensitive to the isospin asymmetry. Furthermore, the Dirac mass $m^*_D$ and the nonrelativistic mass $m^*_{NR}$ which
parametrizes the energy dependence of the single particle spectrum,
are both determined from relativistic Dirac-Brueckner-Hartree-Fock
calculations. The nonrelativistic mass shows a characteristic
peak structure at momenta slightly
above the Fermi momentum $\kf$.
The relativistic Dirac mass shows a proton-neutron mass splitting of $m^*_{D,n} <m^*_{D,p}$ in isospin asymmetric nuclear matter. However, the nonrelativistic mass has a reversed mass splitting $m^*_{NR,n} >m^*_{NR,p}$
which is in agreement with the results from nonrelativistic calculations.
\end{abstract}
\maketitle
\section{Introduction}
A highly discussed topic is the isovector dependence of the nuclear force.
This isovector dependence of the nuclear force can be found in the symmetry energy,
the proton-neutron mass splitting,
and the isovector optical potential.

The behavior of the nuclear symmetry energy at high densities
is an important issue in astrophysics,
because the proton fraction  inside a neutron star is strongly dependent on the nuclear symmetry energy.
Therefore, a stiff nuclear symmetry energy leads to a relative proton-rich neutron star,
whereas a soft one results in a neutron star with only a very small proton fraction.
The proton-richness of a neutron star has consequences for the chemical composition
and cooling mechanism of protoneutron stars~\cite{lattimer91,vandalen00,vandalen03a},
mass-radius correlations~\cite{prakash88,engvik94}, critical densities for kaon condensation
in dense stellar matter~\cite{lee96,kubis99}, and the possibility of
a mixed quark-hadron phase in neutron stars~\cite{kutschera00}.
For example, consider the crucial role of the proton-richness
in the thermal evolution of neutron stars.
In fact, if the proton fraction in the core of a neutron star is above a critical value,
the so-called direct Urca processes can occur~\cite{lattimer91,vandalen00,vandalen03a}.
If they occur, the direct
Urca processes will enhance the neutrino emission and neutron star cooling rate by a
large factor compared to the standard cooling scenario, in which the relatively slow modified Urca and two-body neutrino bremsstrahlung processes play a role~\cite{friman79,vandalen01,yakovlev01,vandalen02,vandalen03b}.

The interest for the isospin dependence of the
nuclear forces at its extremes is of recent date, because data for neutron-rich nuclei were rather scarce in the past. However, the forthcoming new generation of radioactive
beam facilities, e.g. the future GSI facility FAIR in Germany, the Rare Isotope
Accelerator planned in the United States of America or SPIRAL2 at GANIL/France, 
will produce huge amounts of new data
for neutron-rich nuclei.

Currently, the isovector dependence of the nuclear force has been investigated in the heavy ion experiments.
For a recent review see~\cite{baran05}.
The observables in these experiments are the $n/p$ flow~\cite{rizzo04,li04}, isospin tracing~\cite{chen05}, isoscaling of intermediate mass fragments (IMF)~\cite{shetty05}, and $\pi^+/\pi^-$ production~\cite{uma98,gaitanos04}.
Heavy ion reactions have the advantage that they allow to test the nuclear forces at supra-normal densities since in intermediate energy reactions compressions of two to three times nuclear saturation density $n_0$ are reached.
However, the asymmetry of the colliding systems is moderate and therefore the isospin effects on the corresponding observables are generally moderate as well. The interpretation of the various data by transport calculations supports at present a value of the symmetry energy around $E_{sym} \sim 32 \ \textrm{MeV}$ at saturation density with a not too soft increase with density.

However, the theoretical predictions for the isospin dependence of nuclear interactions are still very different.
The symmetry energy in relativistic Dirac-Brueckner-Hartree-Fock (DBHF) calculations is found to be significantly stiffer than
in non-relativistic Brueckner-Hartree-Fock (BHF) approaches~\cite{bombaci91}. At moderate densities the DBHF dependence of $E_{\rm
 sym}$ is qualitatively similar to density dependent relativistic mean-field
 parametrizations using $a_4 = 32-34$ MeV~\cite{vretenar03}. However, the density
 dependence of $E_{\rm sym}$ is generally more complex  than in RMF
theory. In particular at high densities $E_{\rm sym}$ shows a
non-linear and more pronounced increase.
In addition, the present predictions for the isospin dependence of
the effective masses differ
substantially \cite{baran05}. BHF calculations
\cite{zuo99,muether02,hassaneen04,Zuo05}, a nonrelativistic \textit{ab initio} approach,
predict a proton-neutron mass splitting of $m^*_{NR,n} > m^*_{NR,p}$ in
isospin asymmetric nuclear matter.
This prediction stands in contrast to the one from relativistic mean-field (RMF) theory.
When only a
vector isovector $\rho$-meson is included in RMF theory, Dirac phenomenology
predicts equal masses $m^*_{D,n}= m^*_{D,p}$. The inclusion of the
scalar isovector $\delta$-meson, i.e. $\rho+\delta$, in this theory leads even to
$m^*_{D,n} < m^*_{D,p}$ \cite{baran05,liu02}.
The nonrelativistic mass derived from RMF theory shows the same behavior as its
Dirac mass, namely  $m^*_{NR,n} < m^*_{NR,p}$ \cite{baran05}.
The various Skyrme forces give opposite predictions
for the neutron-proton mass splitting and also for
the energy slope of the isovector optical potential.

Relativistic {\it ab initio} calculations which are based
on realistic nucleon-nucleon interactions,
such as the DBHF approach,
are the proper tool
to answer these questions.
Therefore, in the present paper,
which is an extension of the work done in Ref.~\cite{vandalen05},
the DBHF approach based on projection techniques is used
to determine properties of symmetric and asymmetric nuclear matter.
The momentum, density, and isospin dependence of these properties are investigated.
The DBHF results for the symmetry energy are compared to results
from some phenomenological approaches. The application of the DBHF approach allows
one to determine the Dirac mass and the nonrelativistic mass from the same approach.
The results are compared to nonrelativistic BHF and RMF approaches.
In addition, the isovector optical nucleon potential,
which is of importance for transport models in relation
with the collisions of radioactive nuclei, is investigated.

The present paper is organized as follows. In Sec.~\ref{sec:RBA}
we give a short description of the relativistic
DBHF approach. In Sec.~\ref{sec:EM}, we
survey the different definitions and physical concepts of the
effective nucleon mass. In Sec.~\ref{sec:RD}, we present the
results derived from the DBHF approach, which is based on projection techniques, in isospin symmetric and asymmetric nuclear matter and investigate the momentum, density, and isospin dependence of the nucleon effective masses and the optical potentials.
Section~\ref{sec:SC} contains a summary and the
conclusions of our work.
\section{Relativistic Brueckner Approach}
\label{sec:RBA}
In the relativistic Brueckner approach nucleons are dressed
inside nuclear matter as a consequence of their two-body interactions
with the surrounding particles. Starting point is
the in-medium interaction, i.e. the $T$ matrix.
It is treated in the ladder
approximation of the relativistic Bethe-Salpeter (BS) equation
\beqa
T = V + i \int  V Q G G T,
\label{subsec:SM;eq:BS}
\eeqa
where $V$ denotes the bare nucleon-nucleon
interaction and $G$ the intermediate off-shell nucleon.
The Pauli operator Q accounts for the Pauli principle preventing
the scattering to occupied states.
The Green's function $G$ describes the propagation of dressed
nucleons in the medium and fulfills the Dyson equation
\beqa
G=G_0+G_0\Sigma G.
\label{subsec:SM;eq:Dysoneq}
\eeqa
$G_{0}$ denotes the free nucleon propagator, whereas the influence of the
nuclear medium is expressed by the self-energy $\Sigma$.
In the Brueckner formalism  this self-energy $\Sigma$ is determined by summing up the
interactions with all the nucleons inside the Fermi sea $F$ in
Hartree-Fock approximation
\beqa
\Sigma = -i \int\limits_{F} (Tr[G T] - GT ).
\label{subsec:SM;eq:HFselfeq1}
\eeqa
The coupled set of
Eqs.~(\ref{subsec:SM;eq:BS})-(\ref{subsec:SM;eq:HFselfeq1})
represents a self-consistency problem and has to be solved by iteration.
The self-energy consists of scalar $\Sigs$ and vector
$\Sigma^\mu = (\Sigo,  \textbf{k} \,\Sigv) $ components
\beqa
\Sigma(k,\kf)= \Sigs (k,\kf) -\gamma_0 \, \Sigo (k,\kf) +
\bfgamma  \cdot \textbf{k} \,\Sigv (k,\kf).
\label{subsec:SM;eq:self1}
\eeqa

The DBHF approach
is the proper tool
to investigate the properties of nuclear matter, but results from DBHF calculations
are still controversial. These results depend strongly  on approximation schemes and
techniques used to determine the Lorentz
and the isovector structure of the nucleon self-energy.  In the present paper,
the projection technique method is used, which requires the knowledge of
the Lorentz structure of the $T$-matrix in (3). For this purpose the
T-matrix has to be projected onto covariant amplitudes. Hence,
the scalar and vector components of the self-energies
can directly be determined from the projection onto Lorentz
invariant amplitudes. We use the
subtracted $T$-matrix representation scheme for the projection
method described in detail in \cite{gross99,vandalen04b}. Projection techniques are rather
complicated, but are accurate. For example, they have been
used in Refs.~\cite{thm87,sehn97,gross99}.

Another frequently used approach, which is called fit method in the following, was
originally proposed by Brockmann and Machleidt
\cite{brockmann90}. In this approach, one extracts the scalar and vector self-energy
components directly from the single particle potential. Hence,
mean values for the self-energy components are obtained where the
explicit momentum-dependence has already been averaged out. In
symmetric nuclear matter this method is relatively reliable.
However, the extrapolation to asymmetric matter introduces two new
parameters in order to fix the isovector dependencies of the
self-energy components. This makes this procedure
ambiguous~\cite{schiller01}.
\begin{figure}[!h]
\begin{center}
\includegraphics[width=0.9\textwidth] {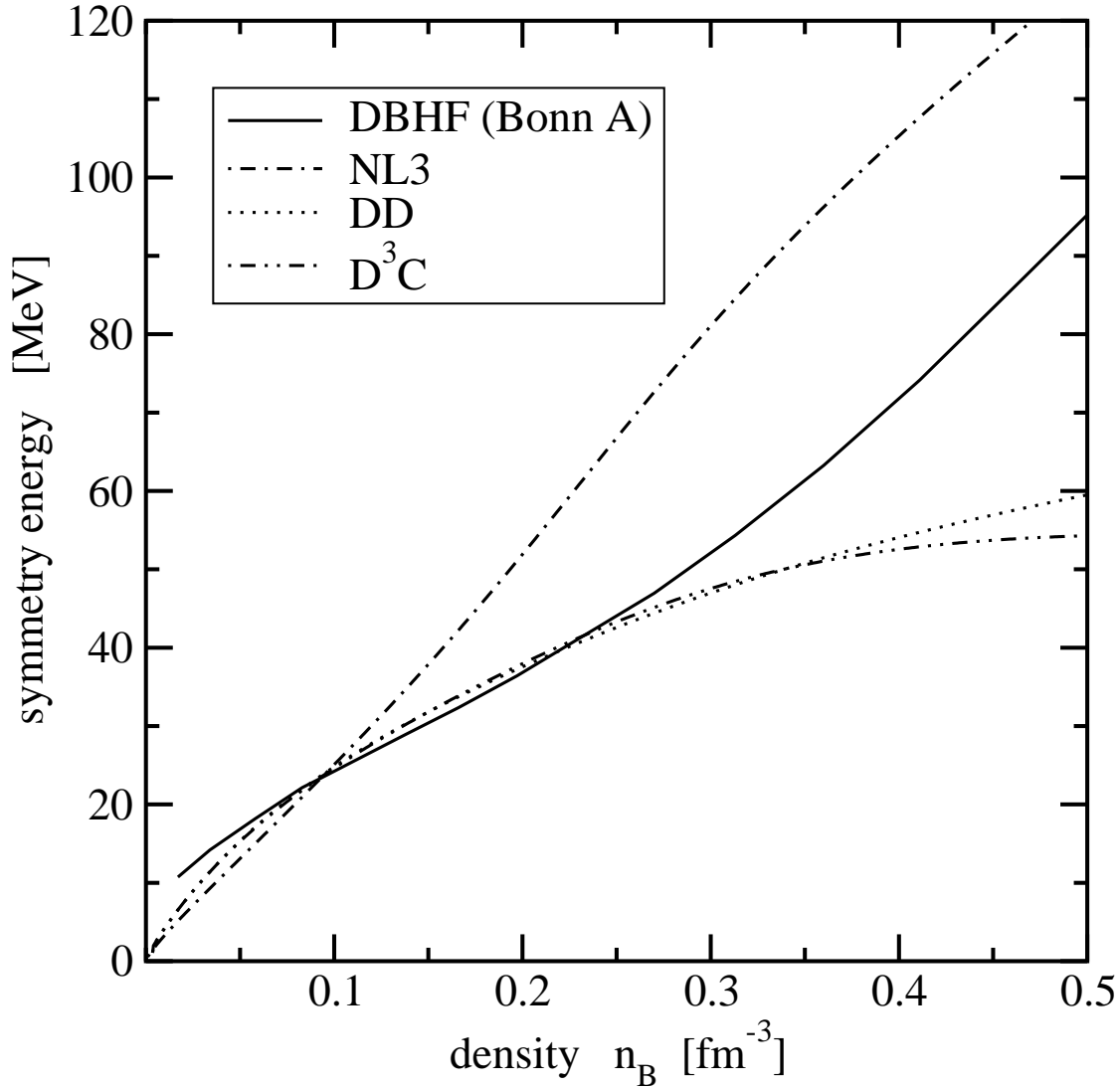}
\caption{Symmetry energy as a function of the nucleon density $n_B$.
The DBHF result is compared to various phenomenological RMF models. 
\label{fig:Esym}}
\end{center}
\end{figure}

The quantity which characterizes the isospin dependence of the nuclear equation of state (EoS) is the symmetry energy. The energy functional of nuclear matter can be expanded in terms of the asymmetry parameter $\beta=(n_n-n_p)/n_B$ 
($n_n$ and $n_p$ are the neutron and proton densities, respectively) which leads to a parabolic dependence on $\beta$
\beqa
E(n_B,\beta)=E(n_B)+E_{sym}(n_B) \beta^2 + {\cal O}(\beta^4).
\eeqa
In Fig.~\ref{fig:Esym} the symmetry energy from the DBHF approach
using the Bonn A potential~\cite{vandalen04b}
is compared to the phenomenological models NL3, DD, and D$^3$C. NL3 is a nonlinear  parametrization~\cite{lalazissis97}
that is widely used in RMF calculations.
The DD model is based on a Lagrangian density of standard relativistic mean-field models
with density dependent meson-nucleon coupling vertices. The D$^3$C
model, in addition, introduces
couplings of the meson fields to derivative nucleon densities
in the Lagrangian density~\cite{typel05}.
The NL3 model has the stiffest EoS and
the symmetry energy rises almost linearly with the density.
In contrast, the DD and D$^3$C model exhibit a considerable flattening.
The DBHF results are more
complex  and have a nonlinear increase at high densities.
At high densities the symmetry energy lies  between the stiff NL3 model and the soft DD and D$^3$C models.
It is worth noting that the symmetry energies in the models
are rather similar at a density near 0.1 fm$^{-3}$. In phenomenological models the symmetry energy is constrained by the skin thickness of heavy nuclei which, due to surface effects, seems to fix the symmetry energy at an average density of about 0.1 fm$^{-3}$. That the DBHF result coincides at this density with RMF phenomenology shows that the low density behavior of the microscopic calculation is in agreement with the constraints from finite nuclei.
\section{Effective Masses}
\label{sec:EM}
In the field of nuclear physics,
the introduction of an effective mass is a common concept to characterize the
quasi-particle properties of a particle inside a strongly interacting
medium. A well established fact is that the effective nucleon mass
in nuclear matter or finite nuclei deviates substantially from
its vacuum value \cite{brown63,jeukenne76,mahaux85}.
However, the expression of an effective nucleon mass  has been used to denote
different quantities, which are sometimes even mixed up: the nonrelativistic effective mass $m^*_{NR}$ and the
relativistic Dirac mass $m^*_{D}$. Although these different definitions of the effective mass are
related, they are based on completely different physical concepts.
Hence, one has to be careful when relativistic and
nonrelativistic approaches are compared on the basis of effective
masses. Whereas the nonrelativistic mass  $m^*_{NR}$ can be
determined from both, relativistic as well as nonrelativistic
approaches, the Dirac mass is a genuine relativistic quantity.
Therefore, the definitions of the relativistic Dirac mass
and of the nonrelativistic mass are given below.
\subsection{Dirac Mass}
The relativistic Dirac mass
is defined through the scalar part of the nucleon self-energy in the
Dirac field equation which is absorbed into the effective mass
\beqa
m^*_D(k, \kf) = \frac{M + \Re  \Sigma_s(k, \kf)}{1+ \Re \Sigv (k, \kf)}~~,
\label{subsec:SM;eq:dirac}
\eeqa
where $\Sigs$ and $\Sigv$ are, respectively, the scalar part
and the spatial vector part of the nucleon self-energy (\ref{subsec:SM;eq:self1}).
The Dirac mass accounts for medium effects through the scalar part
of the self-energy. The correction through the spatial vector
part of the self-energy is generally small \cite{thm87,gross99,vandalen04b}.
Furthermore, the Dirac mass is a smooth function of the momentum.
\subsection{Nonrelativistic Mass}
The effective nonrelativistic mass, which is usually considered in order to characterize the
quasi-particle properties of the nucleon within nonrelativistic frameworks,
is defined as
\beqa
m^*_{NR} = |\veck| [dE/d|\veck|]^{-1}~~,
\label{Landau1}
\eeqa
where $E$ is the quasi-particle's energy and $\veck$ its momentum. When
evaluated at $k=k_F$, Eq.(\ref{Landau1}) yields the Landau mass
$m^*_{L}=M (1+f_1/3)$
related to the $f_1$ Landau parameter of a Fermi liquid \cite{baran05,jaminon89}.
In the quasi-particle approximation, i.e. the zero width limit of the in-medium
spectral function, the quantities $E$ and $m^*_{NR}$ are connected by the dispersion relation
\beqa
E= \frac{\veck^2}{2 M} + \Re U (|\veck|, \kf)~~.
\label{Energy1}
\eeqa
Therefore, equations (\ref{Landau1}) and (\ref{Energy1}) yield the
following expression for the nonrelativistic effective  mass
\beqa
m^*_{NR} = \left[\frac{1}{M}
+  \frac{1}{|\veck|} \frac{d}{ d |\veck|}\Re U \right]^{-1}~~.
\label{mlandau}
\eeqa
In a relativistic framework $m^*_{NR}$ is then obtained from the
corresponding Schroedinger equivalent single particle potential
\beq
U (|{\bf k}|,\kf)
=  \Sigs - \frac{1}{M} \left( E\Sigo - {\bf k}^2\Sigv\right)
    + \frac{\Sigs^2 - \Sigma_{\mu}^2}{2M}  ~.
\label{uopt}
\end{equation}
An alternative would be to derive the effective mass from
Eq. (\ref{Landau1}) via the relativistic single particle
energy 
\beqa
E = (1+\Re \Sigv)\sqrt{ {\bf k}^{2} + m^{*2}_{D}} - \Re  \Sigo.
\eeqa
However, the  single particle energy contains relativistic corrections to the kinetic energy. These kind of corrections should be avoided in a comparison to nonrelativistic approaches.
Hence, the effective mass should be based on the Schroedinger equivalent potential (\ref{uopt})
\cite{jaminon89}.

The nonrelativistic effective mass parametrizes the momentum
dependence of the single particle potential. Therefore, it is a
measure of the nonlocality of the single particle potential $U$.
The nonlocality of $U$ can be due to nonlocalities in space or in
time. The spatial nonlocalities result in a momentum dependence,
whereas nonlocalities in time result in an energy dependence. In
order to separate both effects, one has to distinguish between the
so-called k-mass, which is obtained from Eq. (\ref{mlandau}) at
{\it fixed} energy, and the E-mass, which is given by the
derivative of $U$ with respect to the energy at {\it fixed}
momentum \cite{jaminon89}. Knowledge of the off-shell behavior of
the single particle potential $U$ is needed for a rigorous
distinction between these two masses. The spatial nonlocalities of
$U$ are mainly generated by exchange Fock terms~\cite{muether02,hassaneen04}
and the resulting k-mass is a smooth function of the momentum.
Nonlocalities in time are generated by Brueckner ladder
correlations due to the scattering to intermediate off-shell
states. These correlations are mainly short-range correlations
which generate a strong momentum  dependence with a characteristic
enhancement of the E-mass slightly above the Fermi surface
\cite{mahaux85,jaminon89,muether02,hassaneen04}. The effective nonrelativistic
mass defined by Eqs. (\ref{Landau1}) and (\ref{mlandau}) is given
by the product of k-mass and E-mass \cite{jaminon89}. Thus, it
contains both, nonlocalities in space and time. Therefore, it
should also show a  typical peak structure around $\kf$. This peak
structure reflects - as a model independent result - the increase
of the level density due to the vanishing imaginary part of the
optical potential at $\kf$, which for example is seen in shell model
calculations \cite{jeukenne76,mahaux85,jaminon89}. However, one should
account for correlations beyond mean-field or Hartree-Fock in
order to reproduce this behavior.
\section{Results and Discussion}
\label{sec:RD}
In the following we present the results for the properties of symmetric and asymmetric nuclear matter
obtained from the  DBHF approach based on projection techniques.
The nucleon-nucleon potential used is
Bonn A. However, the presented results and the following discussion do not
strongly depend on the particular choice of the interaction.
\subsection{Symmetric Nuclear Matter}
\begin{figure}[!h]
\begin{center}
\includegraphics[width=0.9\textwidth] {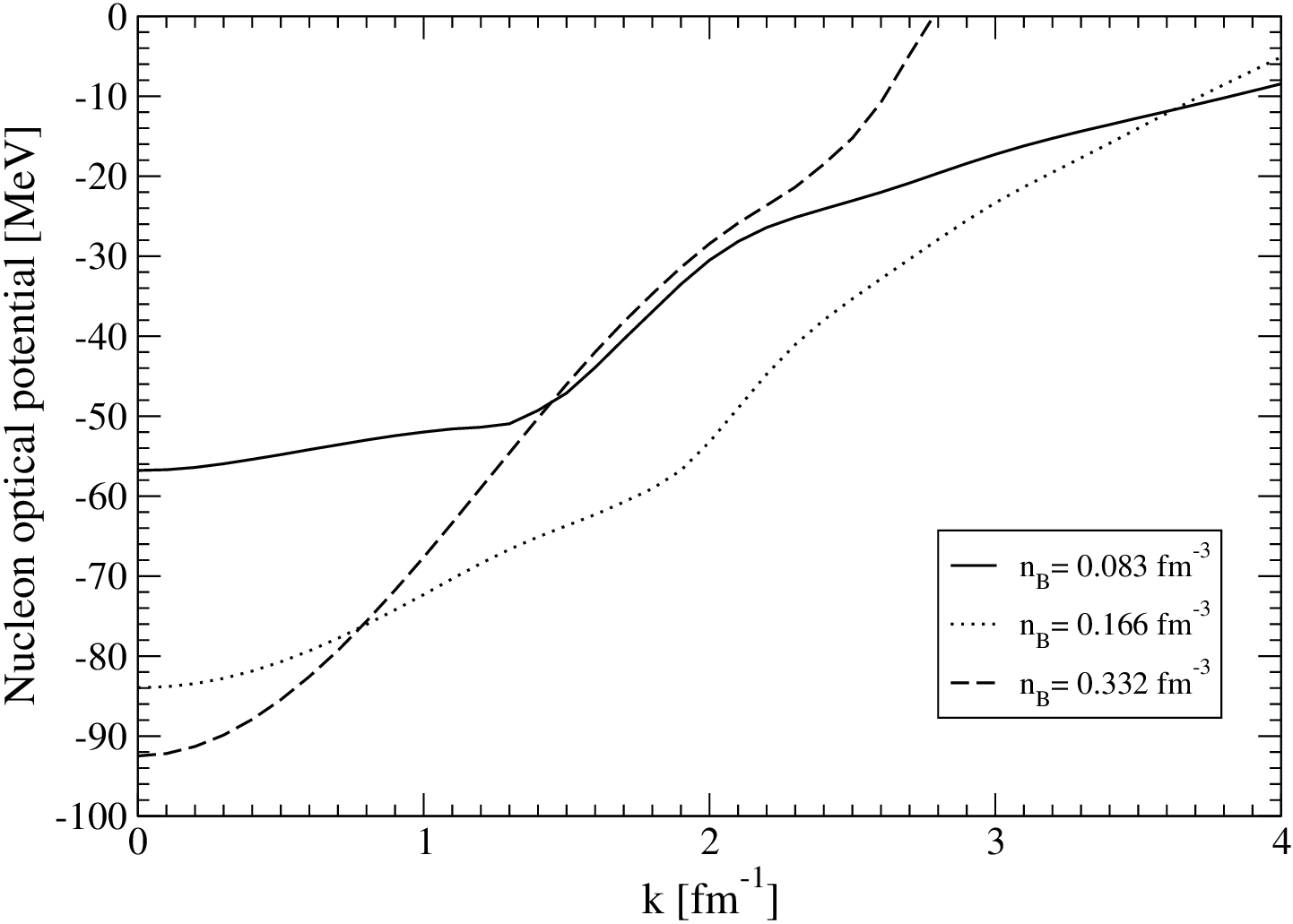}
\caption{The nucleon optical potential in isospin symmetric nuclear matter
as a function of the
momentum $k=|\veck|$ at different densities.
\label{fig:optpot}}
\end{center}
\end{figure}
In Fig.~\ref{fig:optpot} the nucleon optical potential, which is closely related to the
nonrelativistic mass, is plotted
as a function of the momentum $k=|\veck|$ at different Fermi momenta
of $\kf=1.07,~1.35,~{\rm and}~1.7~{\rm fm}^{-1}$ which corresponds to
nuclear densities $n_B = 4 \kf^3 /6\pi^2  = 0.5n_0,~n_0,~{\rm and}~2 n_0$
with $n_0=0.166~{\rm fm}^{-3}$. The depth of the nucleon optical potential at k=0
is larger at higher densities.
Furthermore, the potential increases with momentum at all three densities.
However, the slope of the optical potential is steeper at higher densities.

\begin{figure}[!h]
\begin{center}
\includegraphics[width=0.9\textwidth] {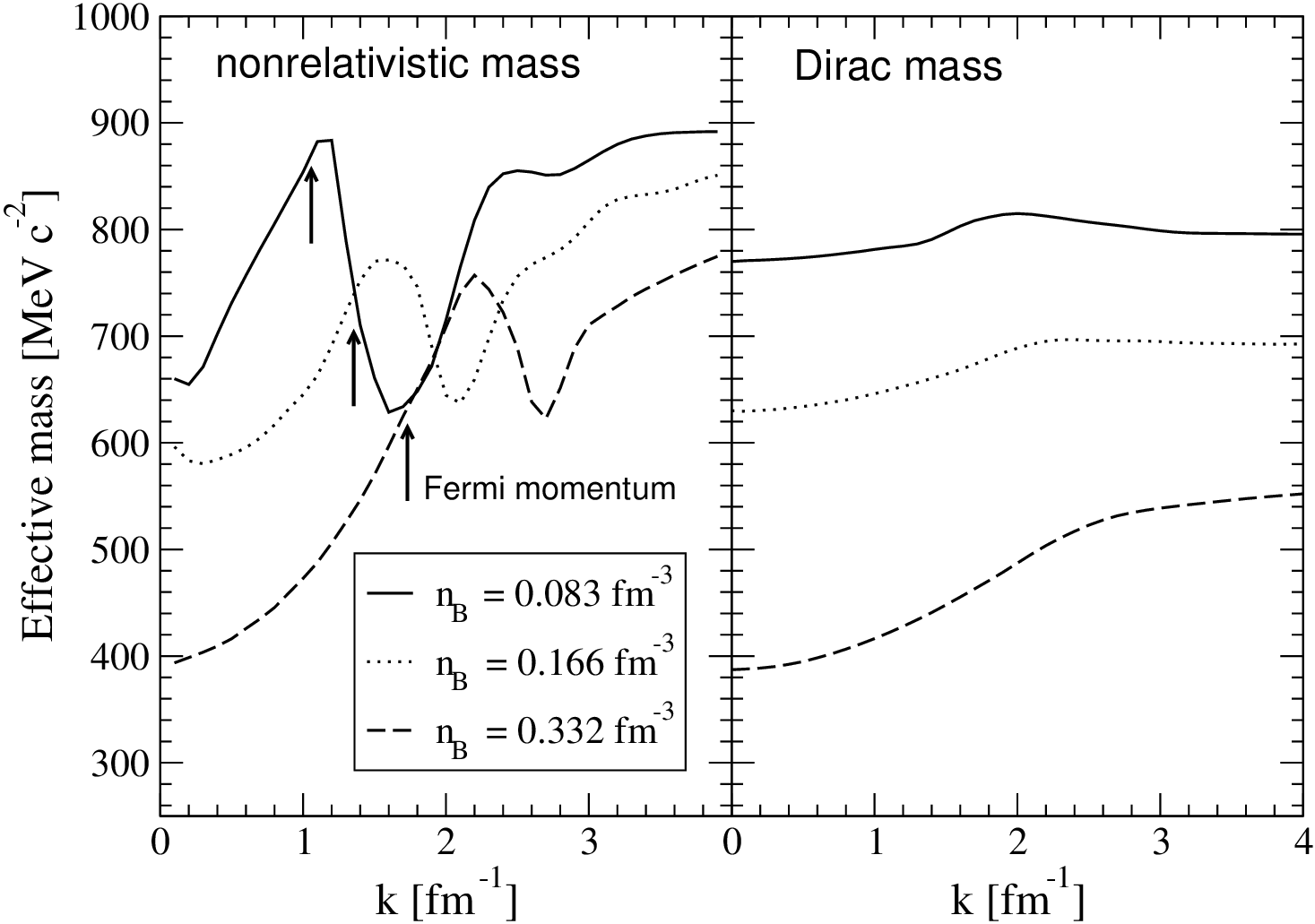}
\caption{The effective mass in isospin symmetric nuclear matter
as a function of the
momentum $k=|\veck|$ at different densities.
\label{fig:sLD}}
\end{center}
\end{figure}
In Fig.~\ref{fig:sLD} the nonrelativistic effective mass
and the Dirac mass are
shown as a function of momentum $k$ at nuclear densities
$n_B = 4 \kf^3/6\pi^2  = 0.5n_0,~n_0,~{\rm and}~2 n_0$.
Both, Dirac and nonrelativistic mass, decrease in average with increasing nuclear
density. The decrease of the nonrelativistic mass
could already be expected on the basis of the slope of the optical potential in Fig.~\ref{fig:optpot}.
The projection method reproduces a pronounced peak of the
nonrelativistic mass slightly above the Fermi momentum  $\kf$,
as it is also seen in nonrelativistic
BHF calculations \cite{jaminon89}.
This peak is shifted to higher momenta and
slightly broadened with increasing density.
On the other hand, the Dirac mass is a smooth function of k with only
a moderate momentum dependence. This behavior is in agreement
with the ``reference spectrum approximation'' used
in the self-consistency scheme of the DBHF approach~\cite{vandalen04b}.

\begin{figure}[!h]
\begin{center}
\includegraphics[width=0.9\textwidth] {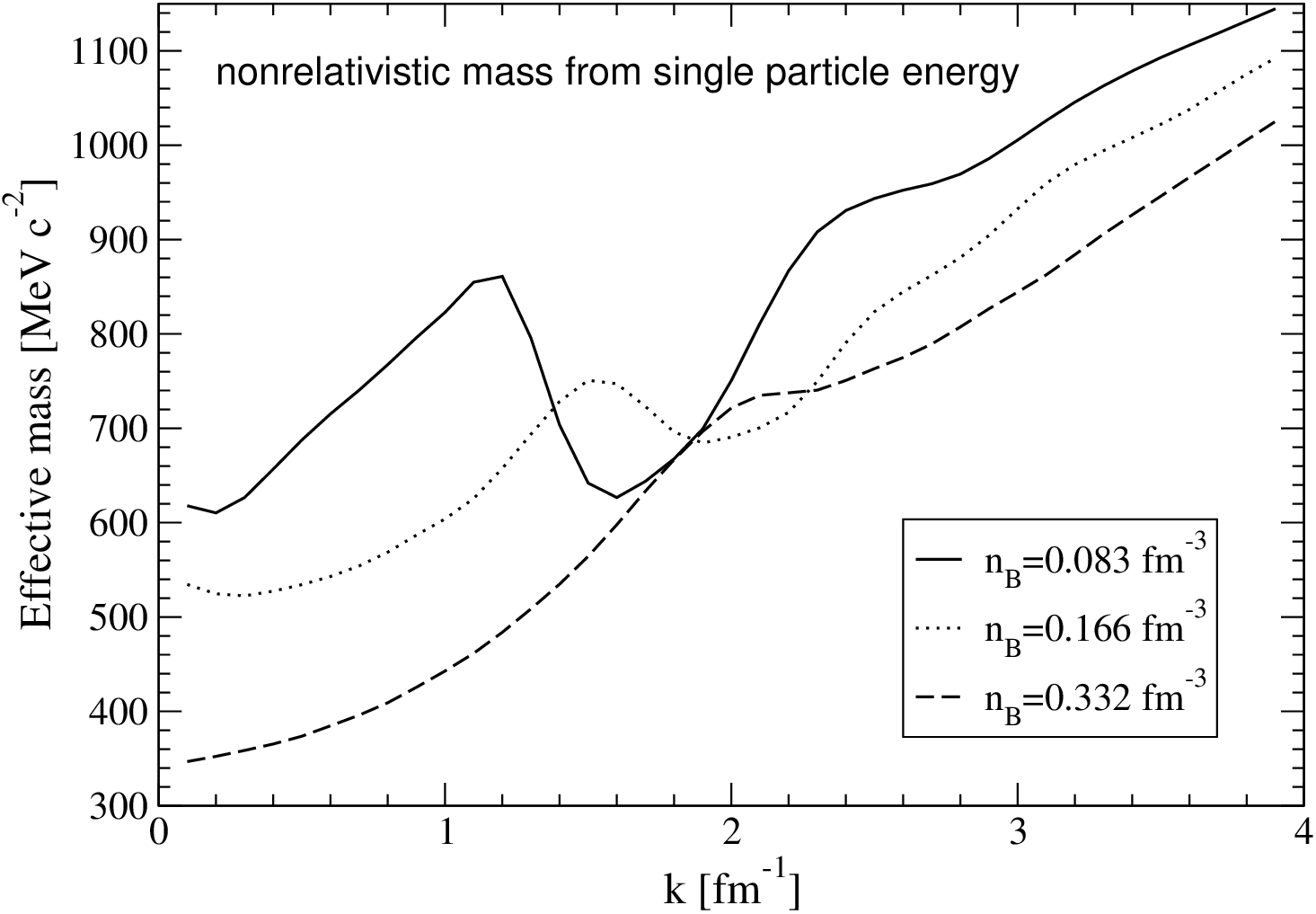}
\caption{The nonrelativistic effective mass in isospin symmetric nuclear matter
extracted from the single particle energy (\ref{Landau1})
as a function of the
momentum $k=|\veck|$ at different densities.
\label{fig:spenergy}}
\end{center}
\end{figure}
The nonrelativistic mass, plotted in Fig.~\ref{fig:spenergy}, is
derived from Eq. (\ref{Landau1}) via the single particle energy
instead of from Eq. (\ref{uopt}) via the potential. The results are very similar to the ones in Fig.~\ref{fig:sLD}.
Again the pronounced peak of the
nonrelativistic mass slightly above the Fermi momentum  $\kf$ is reproduced, although it is
more broadened and as a result it is more a broad bump at high densities.
The important difference is the strong increase of the effective nonrelativistic mass
at high momentum compared to the nonrelativistic mass extracted from the potential.
Relativistic corrections to the kinetic energy
are responsible for this high momentum behavior.
Hence, a comparison to nonrelativistic approaches should
be based on the Schroedinger equivalent potential (\ref{uopt})
\cite{jaminon89}.

Relativistically, the single particle potential and
the corresponding peak structure of the nonrelativistic mass are the result
of subtle cancellation effects of the scalar and vector self-energy components.
Therefore, this requires
a very precise method in order to determine variations of the
self-energies  $\Sigma$ which are small compared to their absolute scale. The
applied projection techniques are the adequate tool for this purpose.
Less precise methods yield only a small enhancement, i.e. a broad bump
around the Fermi momentum $\kf$ \cite{jaminon89,thm87}.
The extraction of mean scalar and vector self-energy components from a fit to
the single particle potential, is not able to resolve such
a structure at all.

\begin{figure}[!h]
\begin{center}
\includegraphics[width=0.9\textwidth] {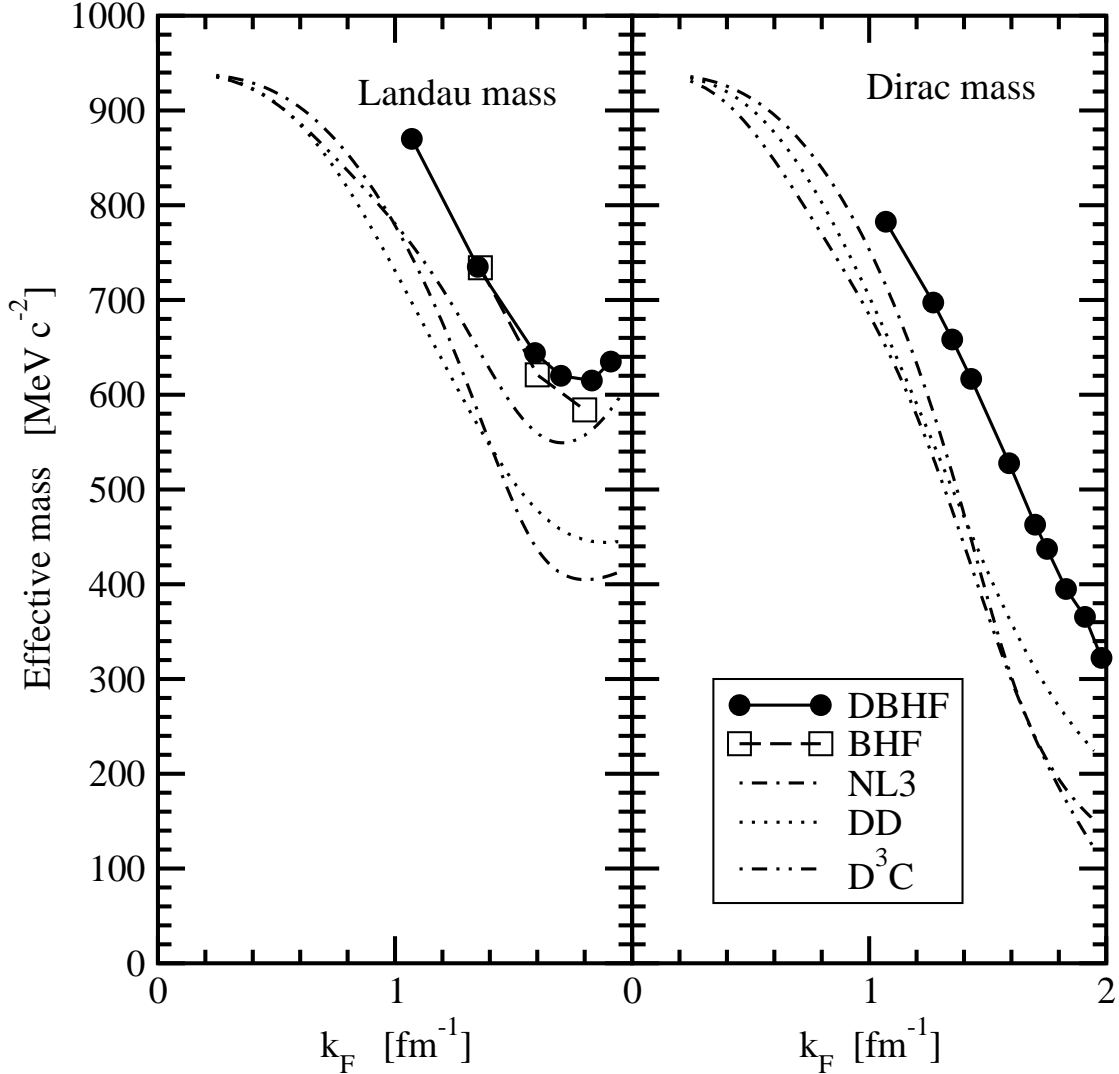}
\caption{The effective mass in isospin symmetric nuclear matter
at $k=|\veck|=\kf$ as a function of the
Fermi momentum $\kf$ for the various models.
\label{fig:mstar}}
\end{center}
\end{figure}
The density dependence of the two
effective masses is compared in Fig.~\ref{fig:mstar}.
Both, the nonrelativistic (Landau)  and the Dirac mass are determined at
$k=|\veck|=\kf$ and shown as a function of $\kf$.
The Dirac mass decreases continously with
increasing Fermi momentum~$\kf$. Initially, the Landau  mass decreases
with increasing Fermi momentum $\kf$ like the Dirac mass.
However, it starts to rise again at high values of the Fermi momentum $\kf$.
In addition, also results from nonrelativistic BHF calculations~\cite{muether},
which are based on the same  Bonn A  interaction, are plotted.
The agreement between the nonrelativistic and the relativistic
Brueckner approach is quite good. This demonstrates that the often discussed difference 
between effective masses obtained in the various approaches is mainly due to different definitions,
i.e. nonrelativistic mass versus Dirac mass. If the same quantity is determined from DBHF and BHF, 
this leads to results which are very close. 
Furthermore, also results from the NL3, DD, and D$^3$C model are shown.
They qualitatively show the same behavior as the Brueckner approaches, i.e.
the Landau mass and the Dirac mass decrease with increasing Fermi momentum. However,
the Landau mass starts to rise again at high values of the Fermi momentum $\kf$.
But quantitatively these masses are lower
compared to the ones in the Brueckner approaches.
\subsection{Asymmetric Nuclear Matter}
\begin{figure}[!h]
\begin{center}
\includegraphics[width=0.9\textwidth] {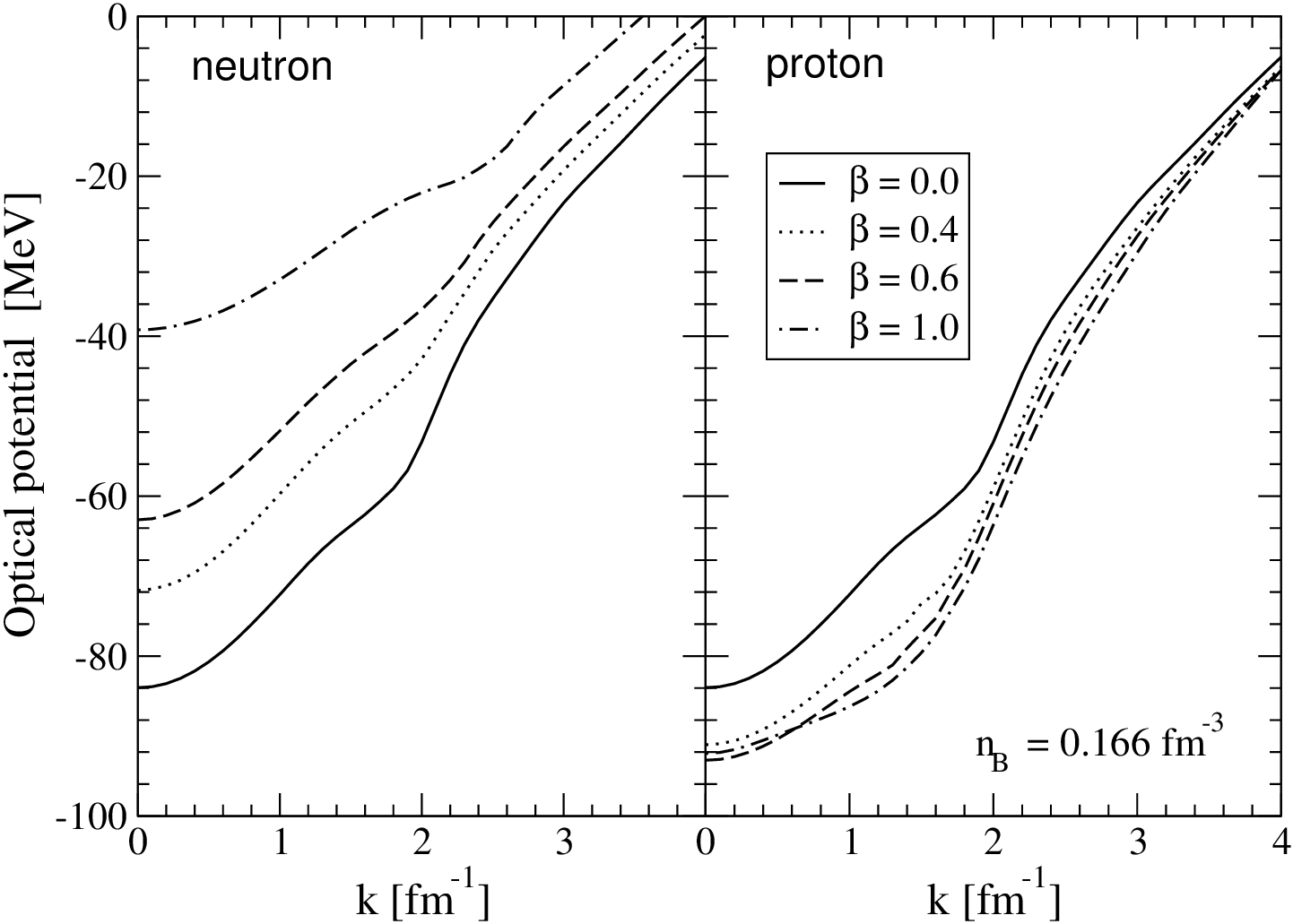}
\caption{The neutron and proton optical potential in isospin asymmetric nuclear matter
as a function of the
momentum $k=|\veck|$ at different densities.
\label{fig:aUopt}}
\end{center}
\end{figure}
In Fig.~\ref{fig:aUopt}
the neutron and proton optical potentials in isospin asymmetric nuclear matter are plotted
as a function of the
momentum $k=|\veck|$ for various values
of the asymmetry parameter $\beta = (n_n - n_p)/n_B$ at fixed nuclear density
$n_B = 0.166 \ \textrm{fm}^{-3}$. The proton optical potential
decreases with increasing asymmetry.
The neutron optical potential, in contrast, shows an opposite
behavior.
In addition, the steepness of the neutron optical potential decreases
with increasing asymmetry parameter $\beta$, whereas the opposite behavior is found
in the proton case.

\begin{figure}[!h]
\begin{center}
\includegraphics[width=0.9\textwidth] {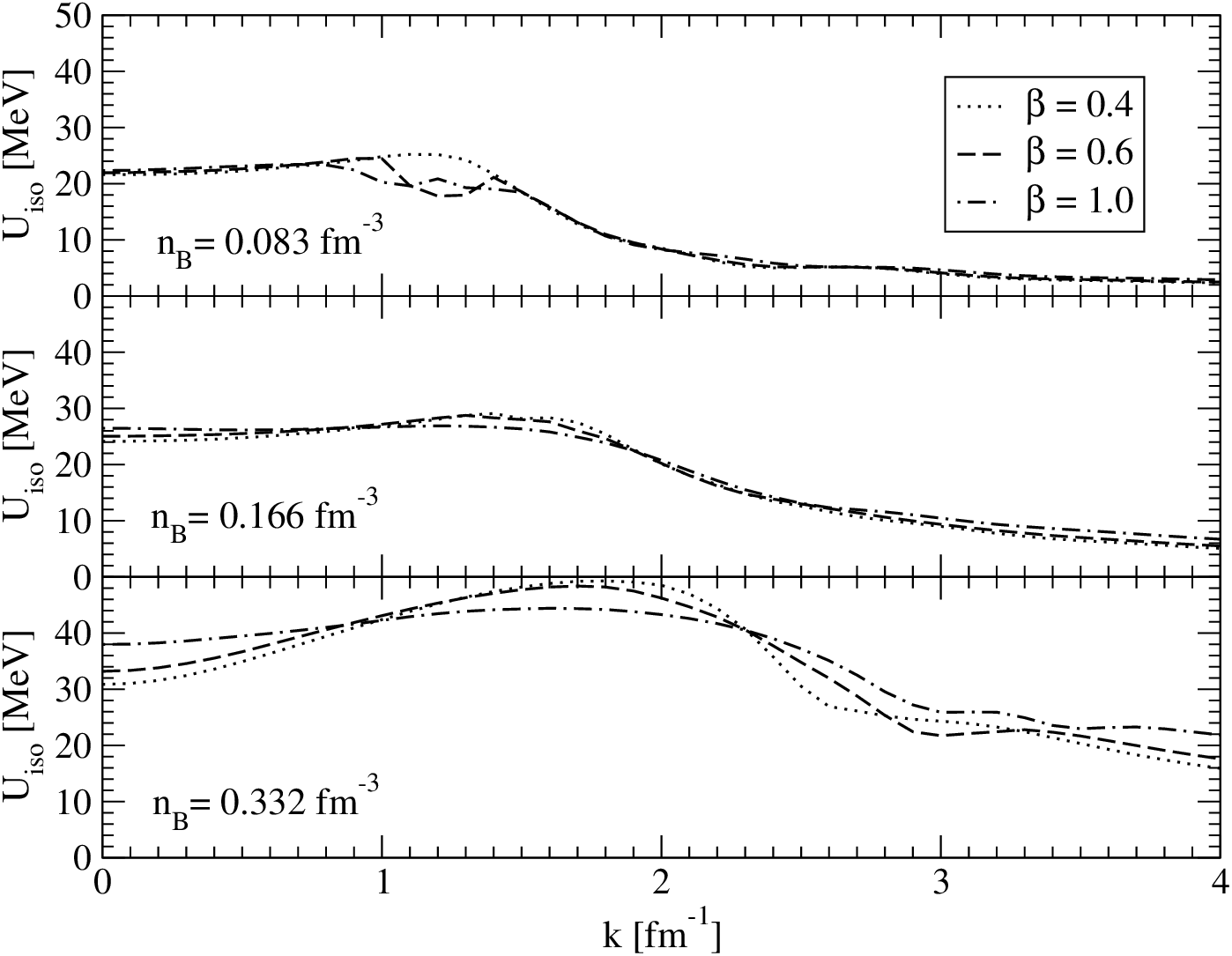}
\caption{The isovector optical potential as a function of momentum
$k$ for three densities and several isospin asymmetries.
\label{fig:isovecUopt}}
\end{center}
\end{figure}
The isovector optical potential 
\begin{equation}
U_{iso}= \frac{U_n - U_p}{2 \beta}
\end{equation}
can be obtained from the neutron and proton optical
potential. In Fig.~\ref{fig:isovecUopt} the isovector optical
potential is displayed as a function of momentum $k$ for three
densities and several isospin asymmetries. It is seen that the
isovector optical  potential depends strongly on density and momentum. The
optical potential in neutron-rich matter stays roughly constant up
to a momentum between 1 to 2 fm$^{-1}$, depending on the density,
and then decreases strongly with increasing momentum. Fig.~\ref{fig:isovecUopt}
shows that the isovector optical potential is almost independent
of the asymmetry parameter $\beta$. The optical isovector
potential at nuclear density $n_B = 0.166 \ \textrm{fm}^{-3}$ at
$k=0$ is in good agreement with the empirical value of 22 - 34
MeV~\cite{li04}.

\begin{figure}[!h]
\begin{center}
\includegraphics[width=0.9\textwidth] {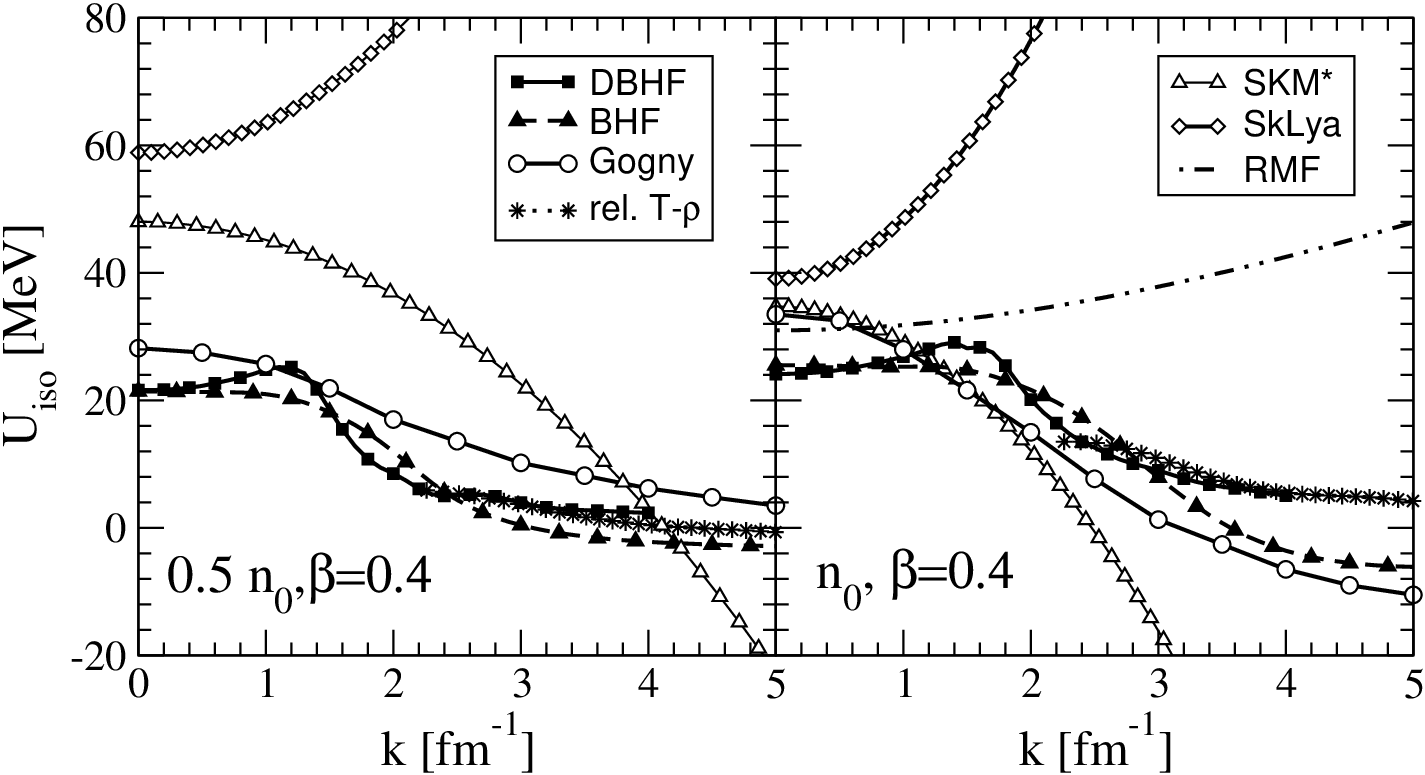}
\caption{The isovector optical potentials from the various models in asymmetric nuclear matter ($\beta=0.4$) as a function of momentum
$k$ at $n=0.5 n_0$ and $n_0$. The isovector optical potential from our DBHF approach  is compared to the ones from the nonrelativistic BHF approach~\cite{Zuo05}, the phenomenological RMF~\cite{gaitanos04}, 
Gogny \cite{gogny02} and Skyrme \cite{skyrme04} forces and from a relativistic $T-\rho$ approximation~\cite{chen05b}. 
\label{fig:isovecUoptb}}
\end{center}
\end{figure}
Fig. (\ref{fig:isovecUoptb}) compares the predictions from our DBHF calculation 
to the nonrelativistic BHF~\cite{Zuo05} 
and to the phenomenological Gogny \cite{gogny02} and Skyrme 
\cite{skyrme04} forces and a 
relativistic $T-\rho$ approximation~\cite{chen05b} based on 
empirical relativistic NN amplitudes \cite{neil83}. 
Our results are in good agreement
with the nonrelativistic BHF results of Ref.~\cite{Zuo05}, except 
for the negative sign of the
potential at high momenta in their work.
In addition, at large momenta our DBHF calculation agrees with the tree-level 
results of~\cite{chen05b}. This is to be expected since 
Pauli blocking of intermediate states 
in the Bethe-Salpeter equation play then a less important 
role. First order medium effects such as a density 
of the effective mass are included in both approaches, in 
 ~\cite{chen05b} within the framework of RMF theory. 

While the dependence of $U_{iso}$ on the asymmetry parameter $\beta$ is 
found to be weak, the predicted energy and density dependences are quite 
different, in particular between the microscopic and the 
phenomenological approaches.
In mean field models, i.e., assuming momentum independent self-energy 
components, the energy dependence of  $U_{iso}$ is linear, i.e. 
quadratic in momentum. Relativisitc mean field models show 
throughout a positive slope \cite{baran05} while Skyrme functionals 
can have positive slopes, e.g. some of the recent Skyrme-Lyon 
parameterisations \cite{skyrme04} (SkLya), or negative ones (SkM$^*$). 
In the former cases this leads to 
a continously increasing optical isovector potential.  SkM$^*$ decreases, 
however, with a much stronger slope than the microscopic approaches 
which tend to saturate at high momenta. Qualitatively such a 
behavior is reproduced by the Gogny force. 
In the DBHF case the decrease is caused by a pronounced explicit momentum 
dependence of the scalar and vector self-energy components.

However, the energy dependence of $U_{iso}$ is very little constrained by data. The old analysis of Lane~\cite{lane62} is consistent with a decreasing potential as predicted by DBHF/BHF, while more recent analyses based on Dirac phenomenology ~\cite{kozack89} come to the opposite conclusions. Certainly more experimental efforts are necessary to clarify this question.

\begin{figure}[!h]
\begin{center}
\includegraphics[width=0.9\textwidth] {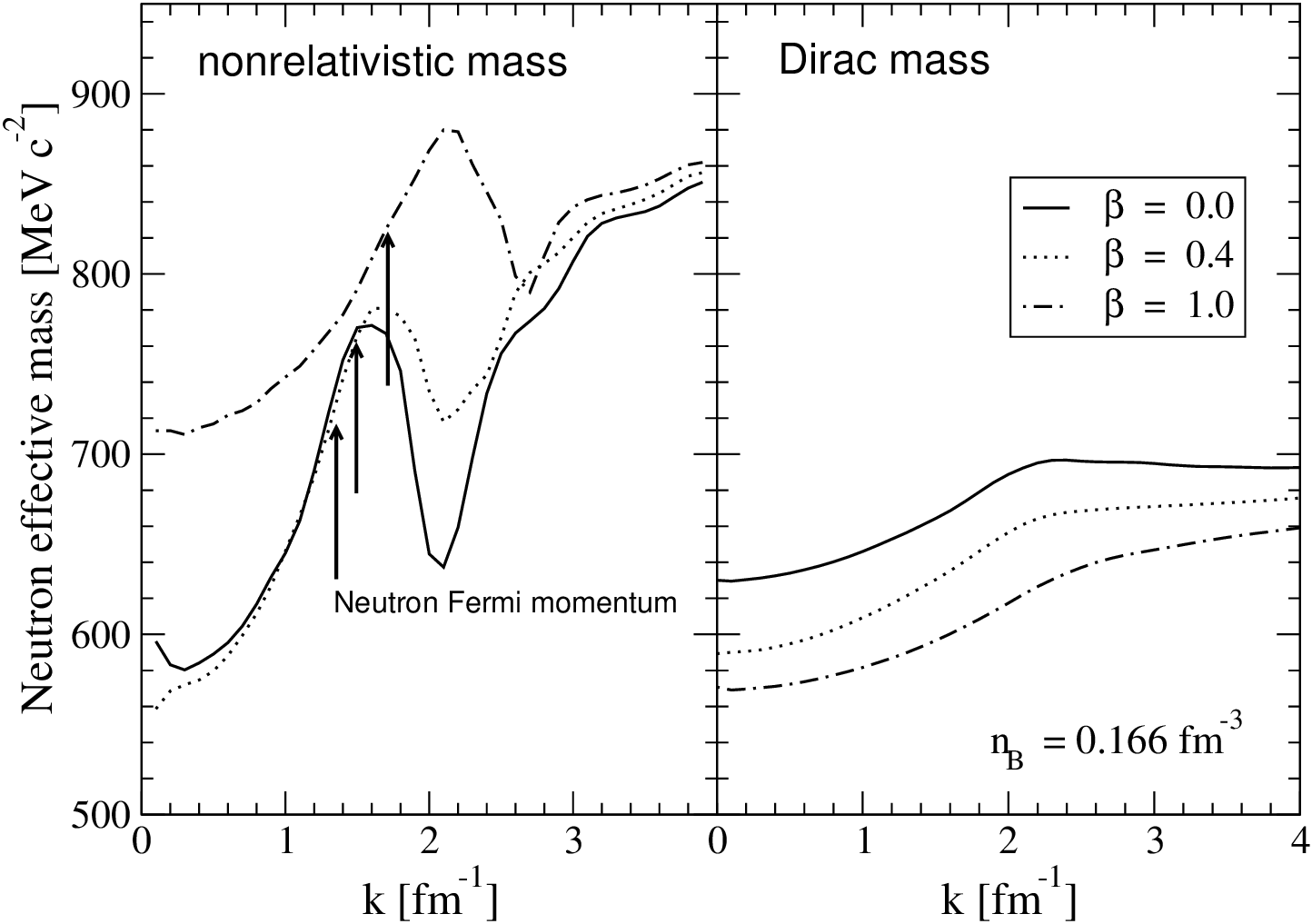}
\caption{Neutron effective mass as a function of the
momentum $k=|\veck|$ for various values
of the asymmetry parameter $\beta$ at fixed nuclear density
$n_B = 0.166 \quad \textrm{fm}^{-3}$.
\label{fig:aLD}}
\end{center}
\end{figure}
In Fig.~\ref{fig:aLD} the neutron nonrelativistic and Dirac mass
are plotted for various values
of the asymmetry parameter $\beta$ at nuclear density
$n_B = 0.166 \quad \textrm{fm}^{-3}$. An increase of
$\beta$ enhances the neutron density and has
for the density of states the same effect as an increase of the density
in symmetric matter. Therefore, a pronounced peak of the
nonrelativistic mass slightly above $\kfn$ is observed.

\begin{figure}[!h]
\begin{center}
\includegraphics[width=0.9\textwidth] {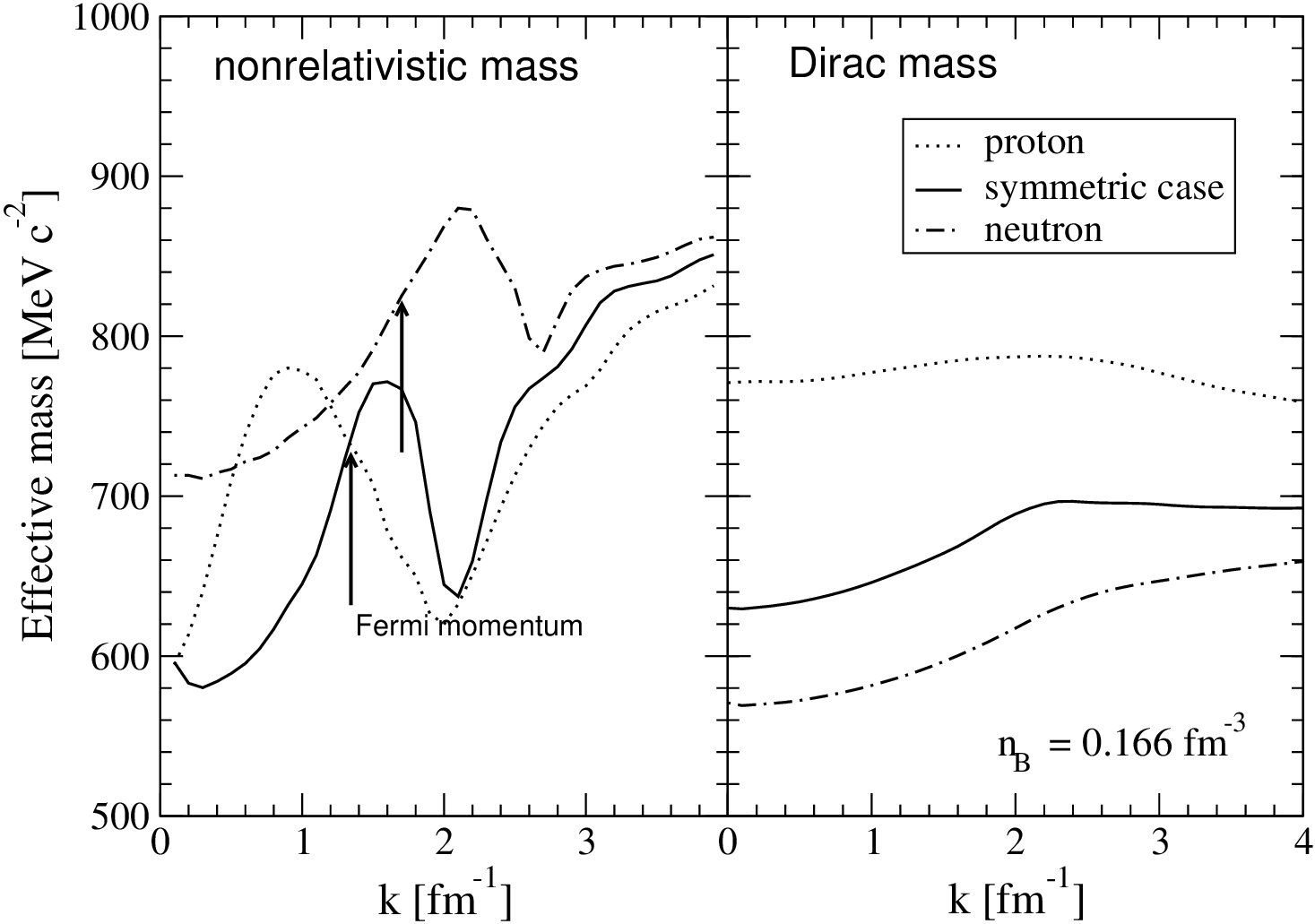}
\caption{Neutron and proton effective mass as a function of the
momentum $k=|\veck|$ for a value
of the asymmetry parameter $\beta=1$ at fixed nuclear density
$n_B = 0.166 \quad \textrm{fm}^{-3}$. In addition,
the effective mass in symmetric nuclear matter is given.
\label{fig:asplit}}
\end{center}
\end{figure}
Another interesting issue is the proton-neutron mass splitting in
isospin asymmetric nuclear matter. In Fig.~\ref{fig:asplit} the
neutron and proton effective mass are compared for $\beta=1$, i.e.
neutron matter. Our DBHF calculations
based on projection techniques predict a mass splitting of
$m^*_{D,n} < m^*_{D,p}$ in isospin asymmetric nuclear
matter. However, the predicted mass splitting based on
the fit method is $m^*_{D,n} > m^*_{D,p}$~\cite{alonso03,sammarruca05}. In the fit method, the
mean values for the self-energy components are obtained where the
explicit momentum-dependence has already been averaged out. In
symmetric nuclear matter this method is relatively reliable.
However, the extrapolation to asymmetric matter introduces two new
parameters in order to fix the isovector dependencies of the
self-energy components. This makes the fit procedure
ambiguous~\cite{schiller01}. Other DBHF calculations
based on projection techniques predict a mass splitting of
$m^*_{D,n} < m^*_{D,p}$ in isospin asymmetric nuclear
matter~\cite{schiller01,dejong98,vandalen04b} in agreement with our results.
Although the relativistic Dirac mass derived from
the DBHF approach based on projection techniques has a proton-neutron mass splitting of
$m^*_{D,n} <m^*_{D,p}$, as can be seen from Fig.~\ref{fig:asplit},
the nonrelativistic mass derived from the DBHF approach shows the
opposite behavior, except around the peak slightly above the proton
Fermi momentum $\kfp$. This opposite behavior to the relativistic
Dirac mass, i.e. $m^*_{NR,n} > m^*_{NR,p}$, is in agreement with
the results from nonrelativistic BHF calculations
\cite{zuo99,muether02,hassaneen04}. This opposite behavior between the Dirac mass splitting and the nonrelativistic mass
splitting is not surprising, since these masses are based on completely different physical concepts.
The nonrelativistic mass parameterizes
the momentum dependence of the single particle potential. 
It is the result of a quadratic parameterization
of the single particle spectrum. On the other hand, the relativistic Dirac mass 
is defined through the scalar part of the nucleon self-energy in the 
Dirac field equation which is absorbed into the effective mass~(\ref{subsec:SM;eq:dirac}). 
\begin{figure}[!h]
\begin{center}
\includegraphics[width=0.9\textwidth] {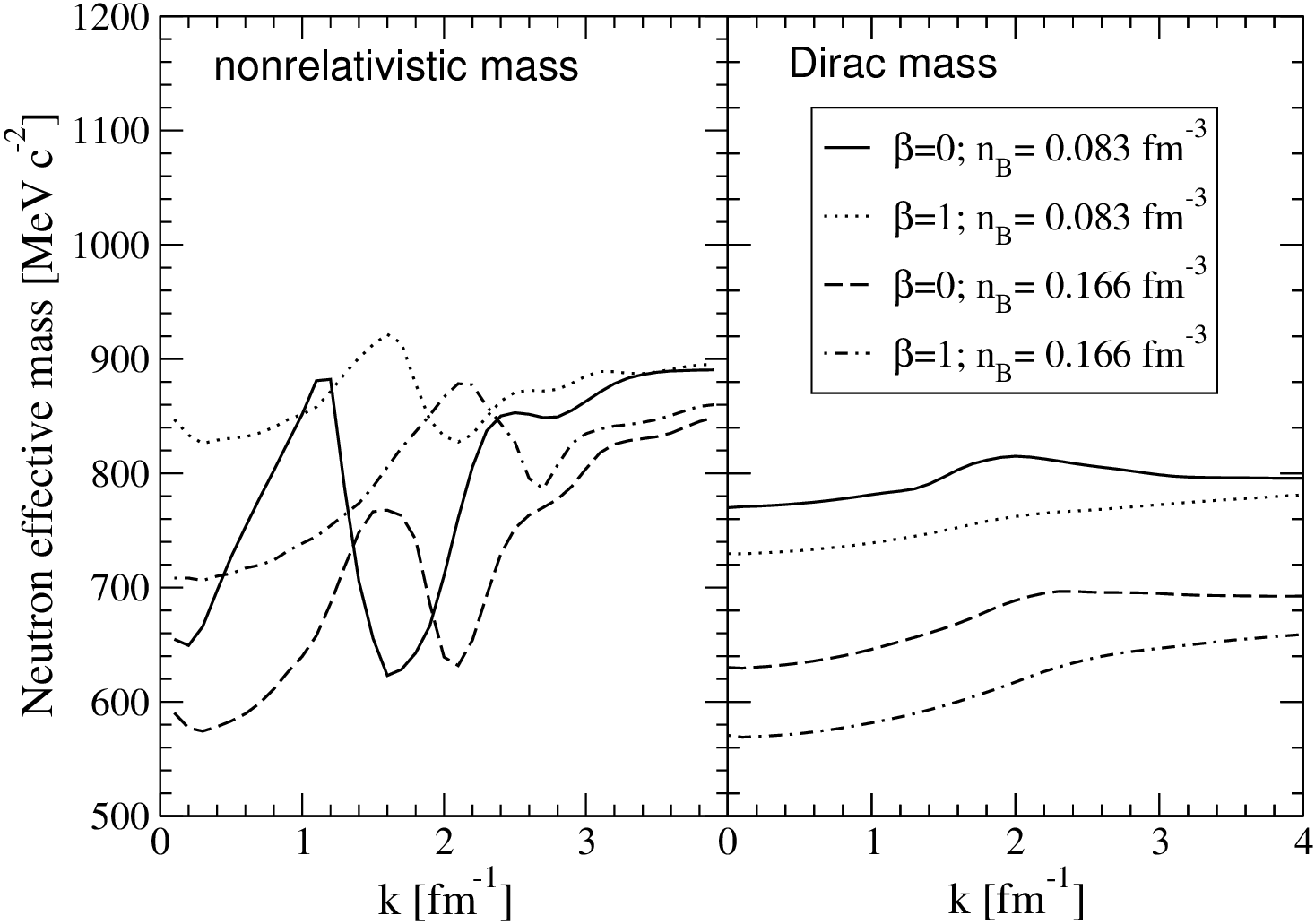}
\caption{Neutron effective mass in symmetric nuclear matter and in pure neutron matter as a function of the
momentum $k=|\veck|$ at different nuclear densities.
\label{fig:adensplit}}
\end{center}
\end{figure}
In Fig.~\ref{fig:adensplit} the neutron nonrelativistic and the neutron Dirac mass in symmetric nuclear matter and in pure neutron matter are plotted at nuclear densities $n_B = 0.083 \quad \textrm{fm}^{-3}$ and 
$n_B = 0.166 \quad \textrm{fm}^{-3}$. 
The difference between the two masses is reduced as the density gets lower, if one excludes the momentum region at the peak structure of the nonrelativistic mass. This peak structure reflects the increase
of the level density due to the vanishing imaginary part of the optical potential at $\kf$. In addition,
with decreasing density the neutron Dirac mass difference in symmetric nuclear and in pure matter gets smaller, i.e. the proton-neutron mass splitting decreases. The same picture can be observed for the nonrelativistic mass, if one does not consider the peak structure of the nonrelativistic mass.

\begin{figure}[!h]
\begin{center}
\includegraphics[width=0.9\textwidth] {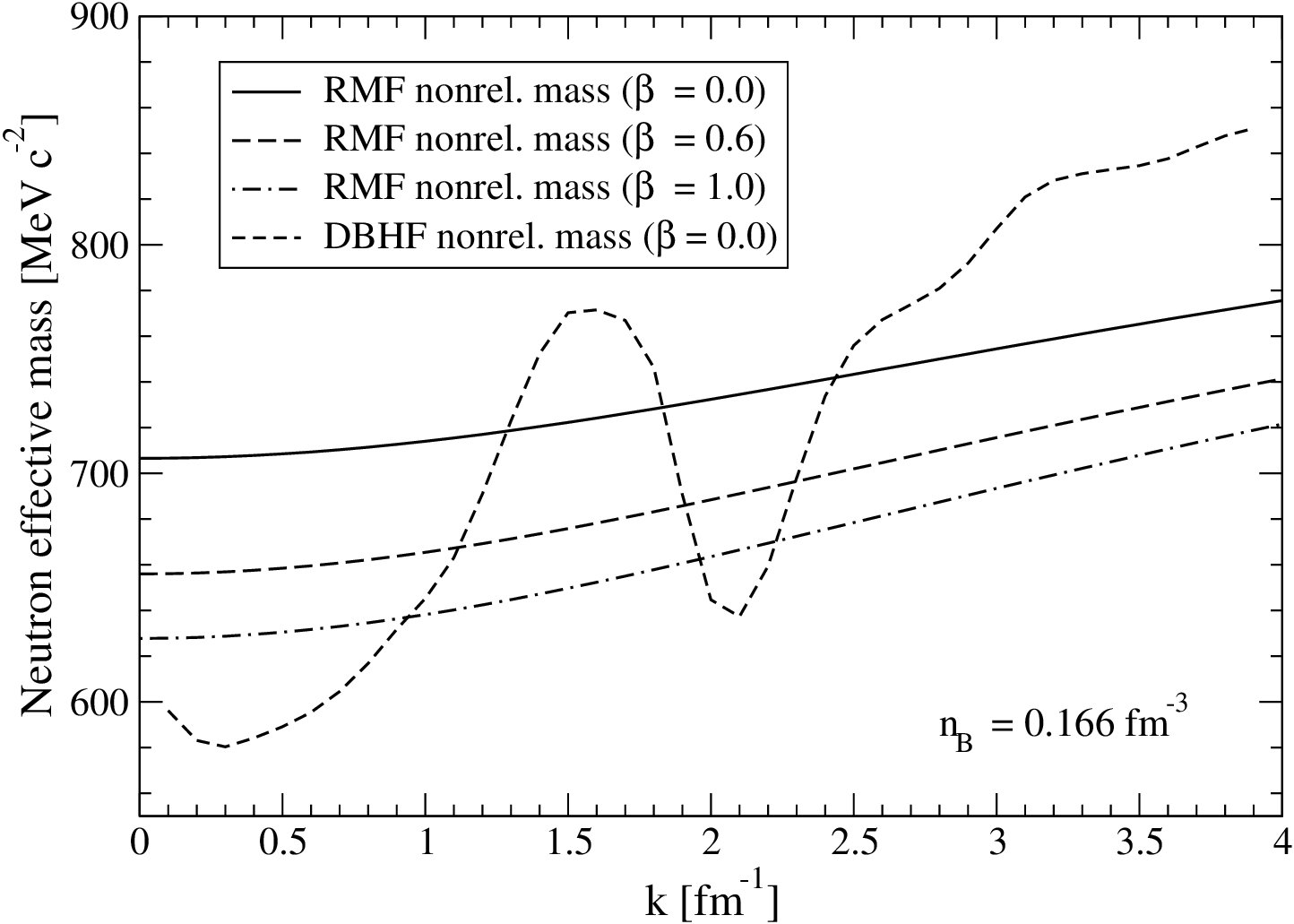}
\caption{Neutron effective mass obtained in the RMF approximation
as a function of the
momentum $k=|\veck|$ at fixed nuclear density $n_B = 0.166 \quad \textrm{fm}^{-3}$.
\label{fig:RMFasymmass}}
\end{center}
\end{figure}
A demonstration of the influence of the
explicit momentum dependence of the DBHF self-energy is shown in Fig.~\ref{fig:RMFasymmass}.
In RMF theory  the relativistic Dirac mass and the vector self-energy
are momentum independent. The nonrelativistic mass is
determined from the RMF approximation to the
single particle potential, i.e. neglecting
the momentum dependence
of the scalar  $\Sigs$ and vector fields $\Sigo$ and $\Sigv$ in
Eqs. (\ref{subsec:SM;eq:dirac}) and   (\ref{uopt}).
The single particle energy is now given by
\beqa
E_{RMF}= (1+\Re \Sigv(\kf)) \sqrt{|\veck|^2+m^{*2}_D(\kf)}+\Re \Sigo(\kf).
\eeqa
In Fig.~\ref{fig:RMFasymmass} this  ``RMF'' nonrelativistic mass is
plotted  for various values
of the asymmetry parameter $\beta$ at fixed nuclear density $n_B = 0.166 \quad \textrm{fm}^{-3}$.
For comparison  the full DBHF nonrelativistic mass for
symmetric nuclear matter is shown as well. Because of the parabolic momentum
dependence of the RMF single particle energy $E_{RMF}$,
the corresponding RMF mass has no bump or
peak structure but is a continuously rising function with momentum.
The nonrelativistic RMF mass at $k=k_F$ corresponds to the RMF Landau mass \cite{jaminon89,matsui81}.
The RMF nonrelativistic mass decreases with increasing asymmetry.
In isospin asymmetric matter RMF theory predicts the same proton-neutron mass splitting for the Dirac
and the nonrelativistic mass, i.e. $m^*_{D,n} < m^*_{D,p}$ and $m^*_{NR,n} < m^*_{NR,p}$.
This behavior is a general feature of the RMF approach \cite{baran05}.
Concerning
the Dirac mass full DBHF theory is in agreement with the prediction of RMF theory.
However, the mass splitting of the nonrelativistic mass is reversed due
to the momentum dependence of the self-energies, which is neglected in RMF theory.
\section{Summary and Conclusions}
\label{sec:SC}
In summary, we present calculations of isospin symmetric and asymmetric nuclear matter in the DBHF approach based on
projection techniques. We compared the momentum, density, and isospin
dependence of the relativistic Dirac mass and the nonrelativistic
mass. Furthermore, we also investigated
these dependencies of the isovector optical potential.
Firstly, the nonrelativistic mass derived from the DBHF approach should
be based on the Schroedinger equivalent potential
(\ref{uopt}) \cite{jaminon89} to be able to compare it
to nonrelativistic approaches. The alternative, to derive it directly from
Eq. (\ref{Landau1}) via the relativistic single particle
energy $E = (1+\Re \Sigv)\sqrt{ {\bf k}^{2} + m^{*2}_{D}} - \Re  \Sigo$,
contains relativistic corrections to the kinetic energy.
Secondly, the nonrelativistic effective mass
shows a characteristic peak structure at momenta slightly
above the Fermi momentum $\kf$ as it is also seen in nonrelativistic BHF calculations, e.g.~\cite{jaminon89}.
This peak structure reflects the increase of the level density at Fermi momentum $k=\kf$.
In contrast, the Dirac mass is a smooth function of k with
a weak momentum dependence.
Thirdly, a strong momentum dependence on both effective masses, the nonrelativistic mass and the Dirac mass,
is observed.
Fourthly, it turns out that the isovector optical potential depends sensitively on density and momentum, but is almost insensitive to the isospin asymmetry.
In addition, the empirical isovector potential extracted from proton-nucleus scattering is well reproduced by our calculation.
Finally, the controversy
between relativistic and nonrelativistic
approaches concerning the proton-neutron
mass splitting in asymmetric nuclear matter has been resolved.
The relativistic Dirac mass shows a proton-neutron mass 
splitting of $m^*_{D,n} <m^*_{D,p}$, in line
with RMF theory. However, the nonrelativistic mass derived from the DBHF approach
has a reversed mass splitting $m^*_{NR,n} >m^*_{NR,p}$
which is in agreement with the results from nonrelativistic BHF calculations.

\begin{acknowledgements}
This work has been supported by
the Deutsche Forschungsgemeinschaft (DFG) under contract no. FA 67/29-1.
\end{acknowledgements}

\end{document}